\documentclass{article}
\usepackage{spconf,amsmath,amssymb,amsfonts,bm,booktabs,multirow,cite,placeins,balance,graphicx}
\usepackage[table]{xcolor}
\usepackage{array,tabularx,pifont}
\usepackage{algorithm,algpseudocode}
\usepackage{caption,subcaption}
\usepackage{enumitem}
\usepackage{tikz}
\usepackage{bm}
\usepackage[flushleft]{threeparttable}
\usepackage[hidelinks]{hyperref}
\definecolor{proposedblue}{RGB}{244,247,250}
\newcolumntype{Y}{>{\centering\arraybackslash}X}
\newcommand{\proposedrow}{\rowcolor{proposedblue}}
\newcommand{\cmark}{\textcolor{green!60!black}{\ding{51}}}
\newcommand{\xmark}{\textcolor{gray}{--}}

\title{Low-Rank Prior-Guided Rank-One Sensing for Efficient CSI Feedback}
\name{Shunpu Tang$^{1}$, Qianqian Yang$^{1}$, Seung-Woo Ko$^{2}$, and Jihong Park$^{3}$}
\address{$^{1}$Zhejiang University, Hangzhou, China,
          $^{2}$Inha University, Incheon, Korea\\
          $^{3}$Singapore University of Technology and Design, Singapore}

\begin{document}
\maketitle

\begin{abstract}
Downlink channel state information (CSI) feedback is essential for frequency-division duplex massive MIMO, yet the feedback overhead grows rapidly with
the numbers of antennas and subcarriers. To reduce this overhead, most deep learning approaches compress CSI by treating it as a generic image, leaving its low-rank multipath structure unexploited. In contrast, model-driven alternatives explicitly embed this structure in iterative recovery, but at the cost of high computational complexity and latency. To overcome these limitations, we propose a low-rank prior-guided (LRP) framework that performs learnable rank-one sensing at the user equipment to compress CSI into low-dimensional codewords, and reconstructs CSI by direct rank-one synthesis at the base station. Both compression and reconstruction are jointly optimized, and fully exploit the low-rank structure of CSI. We further develop DCRNetV2, which preserves LRP as its backbone and uses gated dilated-convolutional residual paths to compensate for finite-rank errors. Experimental results show that LRP outperforms iterative model-based methods with lower complexity, and DCRNetV2 achieves a better accuracy-complexity tradeoff than existing learning-based methods.
\end{abstract}


\section{Introduction}

In frequency-division duplex (FDD) massive MIMO, accurate downlink channel state information (CSI) at the base station (BS) is important for beamforming and interference mitigation \cite{larsson2014massive}. However, because channel reciprocity does not hold, the user equipment (UE) must feed back the CSI over a limited uplink channel \cite{love2008overview}, leading to an extremely high feedback overhead in wideband systems as the numbers of antennas and subcarriers increase.

Compressed sensing reduces this overhead by exploiting angular--delay
sparsity, but their performance can degrade when the assumed sparse representation mismatches practical channels, while iterative recovery introduces non-negligible BS-side computation \cite{rao2014distributed,omp_ref,fista}. Learning-based approaches
avoid explicit inverse optimization by training end-to-end autoencoders to compress and reconstruct CSI \cite{CRNet,DCRNet,TransNet,CLNet, Sun2021Lightweight}. These networks, however, treat CSI primarily as a generic image and learn its physical structure implicitly. In contrast, model-driven designs embed complex signal physical priors into trainable architectures \cite{he2019modeldriven,Guo2022ModelDriven,Ma2025LowComplexity}, nevertheless, they often retain iterative optimization structures through algorithm unfolding.

This raises a simple question: can propagation structure directly drive both CSI compression and reconstruction with low complexity? We answer this by exploiting the low-rank multipath structure of wideband channels. In wideband transmission, each path is essentially a rank-one outer product of separable delay and angular responses. Because the number of dominant paths is limited, the channel matrix is naturally low-rank \cite{Vlachos2019Wideband,Joint_Sparse_and_Low_Rank,zhou2017lowranktensor,shao2022exploiting}. Existing works typically treat this low-rank property using matrix completion, tensor decomposition, or iterative recovery. However, rarely has this path-wise rank-one structure been used as a shared prior to jointly guide CSI compression and direct BS reconstruction.

Motivated by this observation, we formulate CSI feedback as structured sensing and synthesis over rank-one components. Our contributions are threefold.  We propose a low-rank prior-guided (LRP) framework that learns paired delay- and angular-domain sensing vectors at the UE and directly synthesizes CSI from rank-one factors at the BS. Building upon LRP, we further propose DCRNetV2, which retains LRP as its main path and introduces a gated dilated-convolutional architecture to capture residual channel structure beyond the dominant low-rank component. We evaluate LRP and DCRNetV2 on COST 2100 channels, showing that LRP outperforms iterative recovery with lower complexity, and that DCRNetV2 achieves a better accuracy-complexity tradeoff than existing learning-based methods.

\section{System Model}\label{sec:system_model}
We consider an FDD massive MIMO-OFDM downlink with an $N_t$-antenna BS, a single-antenna UE, and $N_c$ subcarriers. The spatial-frequency CSI is represented by $\bm H = [\bm h_1, \ldots, \bm h_{N_c}]^T \in \mathbb C^{N_c \times N_t}$. By applying discrete Fourier transform (DFT) matrices $\bm F_c$ and $\bm F_t$ along with a truncation matrix $\bm S_a$, the UE converts $\bm H$ into the angular-delay domain and retains the first $N_a$ delay bins as $\bm H_a = \bm S_a \bm F_c \bm H \bm F_t^H \in \mathbb C^{N_a \times N_t}$. The UE then encodes this into a compressed vector $\bm z = \mathcal E_{\theta_e}(\bm H_a) \in \mathbb R^M$, from which the BS reconstructs $\hat{\bm H}_a = \mathcal D_{\theta_d}(\bm z)$. The compression ratio is defined as $\eta = M / (2N_aN_t)$, and the network parameters $(\theta_e, \theta_d)$ are trained end-to-end to minimize the mean squared error $\mathbb E[\Vert{}\bm H_a - \hat{\bm H}_a\Vert{}_F^2]$.

\section{Proposed LRP}\label{sec:proposed_lrp}
\subsection{Low-rank multipath prior}
Under the standard far-field geometric channel model with frequency-invariant array responses, the frequency-spatial CSI can be expressed as a superposition of $L$ propagation paths \cite{Vlachos2019Wideband, Joint_Sparse_and_Low_Rank, zhou2017lowranktensor},
\begin{equation}
\bm H=\sum_{\ell=1}^{L}\alpha_\ell
\bm q(\tau_\ell)\bm a^H(\theta_\ell),
\label{eq:geometric_channel}
\end{equation}
where $\alpha_\ell$, $\tau_\ell$, and $\theta_\ell$ denote the complex gain, delay, and angle of the $\ell$th path, respectively. The frequency response $\bm q(\tau_\ell)$ and spatial array response $\bm a(\theta_\ell)$ are separable for each path. Applying the DFTs and retaining the first $N_a$ delay-domain rows preserves this outer-product structure:
\begin{equation}
\bm H_a=\sum_{\ell=1}^{L}\alpha_\ell
(\bm S_a\bm F_c\bm q(\tau_\ell))
(\bm F_t\bm a(\theta_\ell))^H.
\label{eq:angular_delay_channel}
\end{equation}
Since each term in \eqref{eq:angular_delay_channel} is rank one, we have $\operatorname{rank}(\bm H_a)\leq L$. When only a limited number of propagation modes dominate, $\bm H_a$ can therefore be well approximated by $R$ rank-one components,
$
\bm H_a\approx\sum_{r=1}^{R}\bm d_r\bm a_r^H,
$
where $R$ is a model rank controlling the approximation capacity and need not coincide with the physical path count $L$.

\subsection{Rank-one sensing encoder}
Prior matrix-sensing studies have shown that low-rank matrices can be recovered from rank-one projection measurements \cite{cai2015rop,zhong2015efficient}. We parameterize the rank-one sensing operators with learnable weights to adaptively match the underlying channel characteristics. Specifically, we use $R_{\rm enc}$ rank-one sensing branches, where the $r$th branch employs $\bm A_r=\bm u_r\bm v_r^H$ and produces the scalar measurement
\begin{equation}
c_r=\langle\bm A_r,\bm H_a\rangle_F
=\bm u_r^H\bm H_a\bm v_r,
\quad r=1,\ldots,R_{\rm enc},
\label{eq:sensing}
\end{equation}
where $\langle\cdot,\cdot\rangle_F$ denotes the complex Frobenius inner product, and $\bm u_r\in\mathbb C^{N_a}$ and $\bm v_r\in\mathbb C^{N_t}$ are learnable delay- and angular-domain sensing vectors, respectively. Substituting the above low-rank approximation into \eqref{eq:sensing} yields
\begin{equation} 
  c_r \approx \sum_{k=1}^{R} \underbrace{(\bm u_r^H\bm d_k)}_{\text{delay correlation}} \underbrace{(\bm a_k^H\bm v_r)}_{\text{angular correlation}}.
  \label{eq:sensing_expansion} \end{equation}
This can be viewed as an attention-like similarity response: the learnable vectors $(\bm u_r, \bm v_r)$ act as query probes matching the channel's delay-angular keys. The measurement $c_r$ reflects this query-key similarity, producing a strong response only when a path matches both the delay and angular queries.

To obtain the feedback dimension $M$, we stack the sensing outputs into
$
\bm c\in\mathbb C^{R_{\rm enc}},
$
and concatenate their real and imaginary parts as
$
\bar{\bm c}\in\mathbb R^{2R_{\rm enc}}.
$
A lightweight learnable projection then produces the feedback codeword
\begin{equation}
\bm z=\bm W_{\rm enc}\sigma(\bar{\bm c})+\bm b_{\rm enc}
\in\mathbb R^M,
\label{eq:feedback_codeword}
\end{equation}
where $\bm W_{\rm enc}\in\mathbb R^{M\times 2R_{\rm enc}}$ and $\bm b_{\rm enc}\in\mathbb R^M$ are learnable parameters, and $\sigma(\cdot)$ denotes an activation function. Consequently, the LRP encoder embeds the low-rank delay--angular prior directly into the sensing operation, while the subsequent projection adapts the resulting structured measurements to the required feedback dimension.

\subsection{Low-rank multipath reconstruction}
At the BS, the LRP decoder avoids directly generating all $N_aN_t$ complex
CSI coefficients. Instead, it maps $\bm z\in\mathbb R^M$ to $R_{\rm dec}$ pairs of
rank-one factors through a linear projection,
\begin{equation}
  \bm\xi=\bm W_{\rm dec}\bm z+\bm b_{\rm dec}
  \in\mathbb R^{2R_{\rm dec}(N_a+N_t)},
  \label{eq:factors}
\end{equation}
The learnable matrix $\bm W_{\rm dec}$ has $2R_{\rm dec}(N_a+N_t)$ rows and
$M$ columns, and $\bm b_{\rm dec}$ is a bias vector of length
$2R_{\rm dec}(N_a+N_t)$. The output
$\bm\xi$ is partitioned into the real and imaginary parts of
$\hat{\bm d}_r\in\mathbb C^{N_a}$ and
$\hat{\bm a}_r\in\mathbb C^{N_t}$, which denote the delay- and angular-domain
factors of branch $r$, respectively. The CSI is then synthesized as
$
  \hat{\bm H}_a^{\rm dir}=\sum_{r=1}^{R_{\rm dec}}
  \hat{\bm d}_r\hat{\bm a}_r^H.
$
where the complex gains $\hat{\alpha}_r$ are absorbed into the factors. In this way, the branches jointly capture
the dominant low-rank CSI structure without requiring a one-to-one path
correspondence. 
\subsection{End-to-end training}
The structured codec is jointly parameterized by
$\bm\Theta=\{\mathcal U,\mathcal V,\bm W_{\rm enc},\bm b_{\rm enc},
\bm W_{\rm dec},\bm b_{\rm dec}\}$, with
$\mathcal U=\{\bm u_r\}_{r=1}^{R_{\rm enc}}$ and
$\mathcal V=\{\bm v_r\}_{r=1}^{R_{\rm enc}}$. While classical matrix sensing
relies on random projections and iterative recovery without statistical adaptation, we jointly optimize sensing and synthesis by minimizing the expected reconstruction risk:
\begin{equation}
  \min_{\bm\Theta}\,
  \mathbb E_{\bm H_a\sim p_{\rm data}}
  \left[\|\bm H_a-\hat{\bm H}_a\|_F^2\right],
  \label{eq:training}
\end{equation}
via mini-batch stochastic gradient descent. This objective couples sensing with synthesis, ensuring that each measurement preserves the channel structures essential for final CSI reconstruction.

\section{Proposed DCRNetV2}\label{sec:dcrnetv2}
\begin{figure}[t]
  \centering
  \includegraphics[width=1\linewidth]{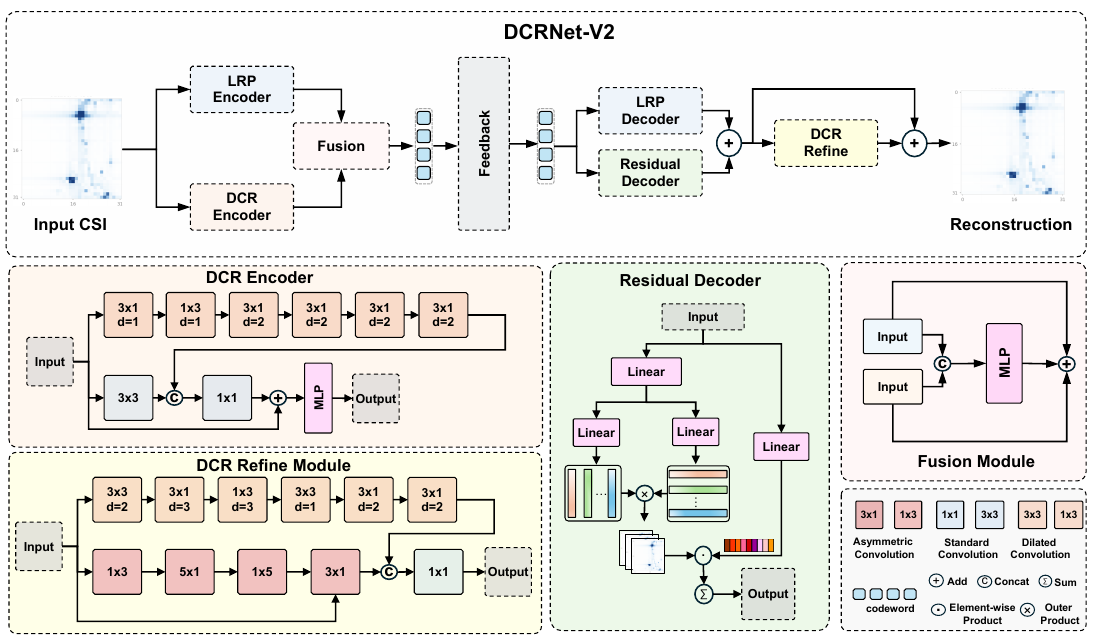}
  \caption{Architecture of DCRNetV2. LRP provides the main low-rank path,
  while DCR modules encode and refine the residual structure.}
  \label{fig:dcrnetv2}
\end{figure}

We then further improve LRP by introducing DCRNetV2, which retains LRP as its main path and adds dilated convolutions blocks from \cite{DCRNet} with design modifications. The overall architecture is shown in Fig.~\ref{fig:dcrnetv2}.

\subsection{Residual encoder and fusion}
At the UE, the angular-delay CSI $\bm H_a$ is processed by two parallel branches. The LRP encoder produces $\bm z_{\rm LRP}\in\mathbb R^M$ from global rank-one measurements. In parallel, a DCR block $f_{\rm DCR}(\cdot)$ extracts local residual features $\bm z_{\rm DCR}=\gamma_e f_{\rm DCR}(\bm H_a)$. Following \cite{DCRNet}, the DCR block uses asymmetric and dilated convolutions: $1\!\times\!k$ and $k\!\times\!1$ kernels separately capture angular and delay correlations, while dilation expands the receptive field to model spatial dependencies of multipath clusters. The auxiliary and fused codewords can be expressed as
\begin{align}
  \bm z&=\bm z_{\rm LRP}+\bm z_{\rm DCR}
  +g_{\rm fuse}([\bm z_{\rm LRP};\bm z_{\rm DCR}]).
  \label{eq:dcr_encoder}
\end{align}
where $\gamma_e$ is a learnable scalar that
controls the DCR contribution, and $[\cdot;\cdot]$ denotes concatenation.
The two-layer MLP $g_{\rm fuse}:\mathbb R^{2M}\rightarrow\mathbb R^M$ learns
the residual interaction between the branches, while the additive paths
preserve their direct information. Initializing $\gamma_e$ and the output
layer of $g_{\rm fuse}$ near zero makes the initial codeword close to
$\bm z_{\rm LRP}$ and allows the residual contribution to grow during
training.
\begin{table}[t!]
  \centering
  \caption{Performance comparison of LRP with classical and model-based methods on COST 2100.}
  \label{tab:main_results}
  \scriptsize
  \setlength{\tabcolsep}{1.6pt}
  \renewcommand{\arraystretch}{1.03}
  \resizebox{\columnwidth}{!}{%
  \begin{tabular}{lrrrrrr}
    \toprule
    & \multicolumn{3}{c}{$\eta=1/4$}
    & \multicolumn{3}{c}{$\eta=1/32$}\\
    \cmidrule(lr){2-4}\cmidrule(lr){5-7}
    Method & FLOPs & Indoor & Outdoor & FLOPs & Indoor & Outdoor\\
    \midrule
    FISTA & 42.99M & $-10.46$ & $-6.35$ & 5.37M & $-1.11$ & $-0.35$\\
    MS4L2O & -- & $-5.96$ & -- & 5.38M & $-1.62$ & --\\
    ISTA & -- & $-4.27$ & -- & 2.75M & $-0.52$ & --\\
    \midrule
   LRP ($R_{\rm enc}=128$, $R_{\rm dec}=4$)
      & \textbf{0.68M} & $-13.18$ & $-4.83$
      & \textbf{0.34M} & $-8.58$ & $-2.40$\\
    LRP ($R_{\rm enc}=512$, $R_{\rm dec}=4$)
      & \underline{1.88M} & $-14.11$ & $-5.96$
      & \underline{1.20M} & $-8.61$ & $-2.64$\\
    LRP ($R_{\rm enc}=512$, $R_{\rm dec}=8$)
      & 2.16M & $\underline{-17.94}$ & $\underline{-8.33}$
      & 1.25M & $\underline{-8.62}$ & $\underline{-2.70}$\\
    LRP ($R_{\rm enc}=512$, $R_{\rm dec}=16$)
      & 2.72M & $\mathbf{-21.62}$ & $\mathbf{-9.04}$
      & 1.34M & $\mathbf{-8.67}$ & $\mathbf{-2.72}$\\
    \bottomrule
  \end{tabular}
  }
\end{table}

\begin{table*}[t]
  \centering
  \caption{Lightweight DL comparison on COST 2100 (NMSE in dB). Best and second-best results are \textbf{bold} and \underline{underlined}.}
  \label{tab:neural_results}
  \scriptsize
  \setlength{\tabcolsep}{2.5pt}
  \renewcommand{\arraystretch}{0.94}
  \begin{tabular}{@{}l*{12}{r}@{}}
    \toprule
    \multirow{2}{*}{Method}
    & \multicolumn{3}{c}{$\eta=1/4$}
    & \multicolumn{3}{c}{$\eta=1/8$}
    & \multicolumn{3}{c}{$\eta=1/16$}
    & \multicolumn{3}{c}{$\eta=1/32$}\\
    \cmidrule(lr){2-4}\cmidrule(lr){5-7}\cmidrule(lr){8-10}\cmidrule(lr){11-13}
    & FLOPs & Indoor & Outdoor & FLOPs & Indoor & Outdoor
    & FLOPs & Indoor & Outdoor & FLOPs & Indoor & Outdoor\\
    \midrule
    CsiNet \cite{CsiNet}
      & 5.41M & $-17.36$ & $-8.75$ & 4.37M & $-12.70$ & $-7.65$
      & 3.84M & $-8.65$ & $-4.51$ & 3.58M & $-6.24$ & $-2.81$\\
    CRNet \cite{CRNet}
      & 5.12M & $-26.99$ & $-12.70$ & 4.07M & $-16.01$ & $-8.04$
      & 3.55M & $-11.35$ & $-5.44$ & 3.28M & $-8.93$ & $-3.51$\\
    CLNet \cite{CLNet}
      & 4.05M & $-29.16$ & $\underline{-12.88}$ & 3.01M & $-15.60$ & $\underline{-8.29}$
      & 2.48M & $-11.15$ & $-5.56$ & 2.22M & $-9.25$ & $\underline{-3.75}$\\
    ACRNet-$1\times$ \cite{ACRNet}
      & 4.64M & $-27.16$ & $-10.71$ & 3.60M & $-15.34$ & $-7.85$
      & 3.07M & $-10.36$ & $-5.19$ & 2.81M & $-8.60$ & $-3.31$\\
    DCRNet-$1\times$ \cite{DCRNet}
      & 4.01M & $-28.04$ & $-12.58$ & 2.96M & $-16.26$ & $-7.95$
      & 2.44M & $-11.74$ & $\underline{-5.60}$ & 2.18M & $-9.05$ & $-3.47$\\
    \midrule
    DCRNetV2-mini
      & $\mathbf{2.83}$M & $-31.00$ & $-10.83$ & $\mathbf{1.87}$M & $-19.23$ & $-7.20$
      & $\mathbf{1.39}$M & $-14.18$ & $-4.83$ & $\mathbf{1.15}$M & $\underline{-9.55}$ & $-3.14$\\
    DCRNetV2-small
      & $\underline{4.00}$M & $-31.12$ & $-11.44$ & $\underline{2.67}$M & $\mathbf{-20.51}$ & $-7.70$
      & $\underline{2.05}$M & $\underline{-14.64}$ & $-5.17$ & $\underline{1.74}$M & $-9.16$ & $-3.31$\\
    DCRNetV2-base
      & 5.23M & $\underline{-32.75}$ & $-11.89$ & 3.88M & $-20.07$ & $-7.83$
      & 3.20M & $\mathbf{-14.73}$ & $-5.51$ & 2.86M & $-9.35$ & $-3.56$\\
    DCRNetV2-unified
      & 6.59M & $\mathbf{-32.84}$ & $\mathbf{-12.99}$ & 4.64M & $\underline{-20.36}$ & $\mathbf{-8.65}$
      & 3.66M & $-14.62$ & $\mathbf{-5.85}$ & 3.17M & $\mathbf{-9.69}$ & $\mathbf{-3.79}$\\
    \bottomrule
  \end{tabular}
\end{table*}

\subsection{Hybrid decoder and refinement}
At the BS, the LRP decoder first reconstructs the direct estimate $\hat{\bm H}_a^{\rm dir}$. Empirically, increasing $R_{\rm dec}$ monotonically improves performance as it captures weak multipath components and residual channel energy. However, scaling $R_{\rm dec}$ directly inflates the projection parameter overhead. To enlarge the effective rank with minimal cost, a residual decoder maps $\bm z$ through $\bm W_{\rm res}\in\mathbb R^{R_e d\times M}$ and reshapes the output into $\bm F\in\mathbb R^{R_e\times d}$, where $R_e$ is the number of additional rank-one components, $d$ is the bottleneck dimension, and $\bm F_r\in\mathbb R^d$ is the $r$-th row of $\bm F$. Shared matrices $\bm W_d\in\mathbb R^{2N_a\times d}$ and $\bm W_a\in\mathbb R^{2N_t\times d}$ map $\bm F_r$ to the real and imaginary parts of $\hat{\bm d}^{e}_r\in\mathbb C^{N_a}$ and $\hat{\bm a}^{e}_r\in\mathbb C^{N_t}$, respectively. The direct and residual paths are merged as
\begin{equation}
  \hat{\bm H}_a^{\rm hyb}=\hat{\bm H}_a^{\rm dir}
  +\alpha\sum_{r=1}^{R_e}g_r(\bm z)
  \hat{\bm d}^{e}_r(\hat{\bm a}^{e}_r)^H,
  \label{eq:dcr_decoder}
\end{equation}
\noindent where $g_r(\bm z)\in[0,1]$ denotes a sample-adaptive sigmoid gate for the $r$-th residual component, and $\alpha$ is a learnable scaling factor. As with the encoder, the residual decoder is initialized near zero to preserve the initial LRP solution.

To further improve reconstruction, we add a final DCR module $f_{\rm ref}(\cdot)$ to refine the hybrid estimate by predicting a residual correction, resulting in the final output
\begin{equation}
  \hat{\bm H}_a=\hat{\bm H}_a^{\rm hyb}
  +\gamma_{\rm ref}f_{\rm ref}(\hat{\bm H}_a^{\rm hyb}).
  \label{eq:dcr_refine}
\end{equation}
where $\gamma_{\rm ref}$ is a learnable scalar gate. This final refinement step is also widely used in existing CSI feedback networks \cite{CRNet,DCRNet,TransNet} to capture residual structure beyond the low-rank approximation.

\subsection{Configurations}
To balance accuracy and complexity, we configure four DCRNetV2 variants via four key hyperparameters: sensing branches $R_{\rm enc}$, direct synthesis branches $R_{\rm dec}$, residual components $R_e$, and bottleneck width $d$, denoted as $(R_{\rm enc}, R_{\rm dec}, R_e, d)$. Specifically, Mini uses $(128, 10, 8, 8)$ and Unified scales to $(512, 32, 16, 16)$. For simpler architectures, Small $(256, 16, 0, 0)$ and Base $(512, 16, 0, 0)$ omit the factored residual decoder ($R_e = 0, d = 0$).
\section{Experiments}
\subsection{Experimental settings}
Following the standard COST 2100 settings \cite{CsiNet,CsiNetPlus,cost2100}, we evaluate indoor (5.3 GHz) and outdoor (0.3 GHz) scenarios with $(N_t, N_c, N_a) = (32, 1024, 32)$. Networks are trained for 1500 epochs using Adam with MSE loss, a batch size of 200, and an initial learning rate of $2 \times 10^{-3}$. We report the normalized mean squared error (NMSE) on the testset, as well as the number of floating-point operations (FLOPs). Baselines include classical and model-based solvers (ISTA, FISTA \cite{fista}, and MS4L2O \cite{liu2023towards}) as well as learning-based models (CsiNet \cite{CsiNet}, CRNet \cite{CRNet}, CLNet \cite{CLNet}, and ACRNet \cite{ACRNet}, DCRNet-$1\times$ \cite{DCRNet}).

\begin{figure}[t!]
  \centering
  \includegraphics[width=0.495\columnwidth,trim={0 196.65pt 0 0},clip]{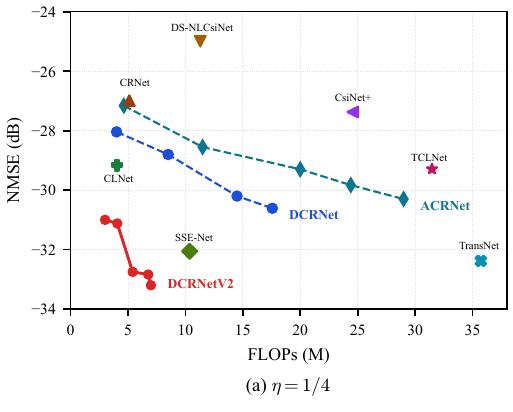}\hfill
  \includegraphics[width=0.495\columnwidth,trim={0 0 0 196.65pt},clip]{figs/flops_vs_nmse.pdf}
  \caption{FLOPs vs.\ indoor NMSE trade-off for representative DL-based CSI
  feedback methods at (a) $\eta=1/4$ and (b) $\eta=1/16$.}
  \label{fig:flops_vs_nmse}
\end{figure}

\subsection{Performance comparison}
Table~\ref{tab:main_results} first compares LRP with classical and model-based algorithms. LRP consistently outperforms iterative methods while incurring lower computational complexity, validating the effectiveness of the multipath low-rank prior. Table~\ref{tab:neural_results} further evaluates DCRNetV2 against existing learning-based methods. DCRNetV2-mini achieves the lowest FLOPs across all compression ratios and surpasses the baselines in most cases. Scaled variants (DCRNetV2-small, -base, and -unified) offer further accuracy gains with moderate complexity increases, with DCRNetV2-unified achieving the best overall performance at most ratios.

Fig.~\ref{fig:flops_vs_nmse} illustrates the accuracy-complexity trade-off among representative DL-based methods. We benchmark DCRNetV2 against the aforementioned baselines at $\eta=1/4$ and $\eta=1/16$, while also incorporating high-complexity benchmarks, including DS-NLCsiNet, CsiNet+, TransNet, and the LLM-powered LLMCsiNet-S \cite{CsiNetPlus,TransNet,LLMCsiNet}. As shown, DCRNetV2 consistently delivers a superior trade-off between reconstruction accuracy and complexity, confirming the architectural efficacy of our design.
\subsection{Ablation studies}

\begin{table}[t]
  \centering
  \caption{Ablation of architectural components on indoor COST 2100 at
  $\eta=1/4$.}
  \label{tab:ablation}
  \footnotesize
  \renewcommand{\arraystretch}{1.04}
  \setlength{\tabcolsep}{2pt}
  \resizebox{0.94\columnwidth}{!}{%
  \begin{tabularx}{\columnwidth}{@{}YYYYYY@{}}
    \toprule
    \multicolumn{2}{c}{\textbf{Enc.}}
    & \multicolumn{2}{c}{\textbf{Dec.}}
    & \multirow{2}{*}{\textbf{NMSE}}
    & \multirow{2}{*}{\textbf{FLOPs}}\\
    \cmidrule(lr){1-2}\cmidrule(lr){3-4}
    \textbf{LRP} & \textbf{DCR} & \textbf{LRP} & \textbf{DCR} & & \\
    \midrule
    \cmark & \xmark & \cmark & \xmark & $-21.62$ & 2.72M\\
    \cmark & \xmark & \cmark & \cmark & $-29.34$ & 3.84M\\
    \cmark & \cmark & \cmark & \xmark & \underline{$-32.00$} & 6.45M\\
    \proposedrow \cmark & \cmark & \cmark & \cmark & $\mathbf{-32.84}$ & 6.59M\\
    \bottomrule
  \end{tabularx}%
  }
\end{table}

Table~\ref{tab:ablation} ablates the two residual paths in DCRNetV2-unified. Incorporating only the decoder- or encoder-side DCR improves NMSE from $-21.62$~dB to $-29.34$~dB and $-32.00$~dB, respectively, while combining both achieves $-32.84$~dB. This highlights the efficacy of the proposed architecture in capturing residual structures and reducing low-rank approximation errors.

\section{Conclusion}
In this paper, we have revisited the problem of CSI feedback as a low-rank matrix sensing and reconstruction problem. We proposed a novel low-rank multipath prior (LRP) framework for CSI feedback. LRP achieves strong performance with low complexity, and can be further extended to DCRNetV2, which combines LRP with dilated convolutional residual paths. Experiments on COST 2100 verify the effectiveness of the proposed methods.

\bibliographystyle{IEEEbib}
\bibliography{references}

\end{document}